\documentstyle[nato,numreferences]{crckapb}

\begin{opening}
\title{QUANTUM INFORMATION AND QUANTUM BLACK HOLES}

\author{JACOB D. BEKENSTEIN}
\institute{The Racah Institute of Physics, Hebrew University
of Jerusalem\\ Givat Ram, Jerusalem 91904, Israel}

\end{opening}

\runningtitle{}

\begin{document}



\section{Limitations on Quantum Information from Black Holes}
\label{sec:BHS}

\subsection{Introduction}
\label{I2} 

A naive view of physics---classical or
quantum---would have us believe that an infinite amount of
information can be contained in a finite 3-D space.  After
all, matter can have an infinity of classical configurations
there, and quantum fields have an infinity of modes in that
region.  't Hooft and Susskind's holographic principle
\cite{thooft,susskind} has shattered this popular view: it
holds that insofar as the information required to describe
them is concerned, physical systems are inherently
two--dimensional in space.  In particular, the information or
entropy (see Sec.~\ref{sec:I&E} below for the relation) 
in an isolated system is expected to be bounded
from above by one quarter the area of a circumscribing
surface expressed in Planck units (holographic bound):
\begin{equation} 
S\leq A(4\hbar)^{-1}
\label{holographic} 
\end{equation} 
Except as otherwise stated, I assume units
with $G=c=1$.   't Hooft's original statement of the
holographic principle \cite{thooft} was elucidated by Susskind
\cite{susskind} who showed that the holographic bound is
required  by the generalized second law (GSL) \cite{bek72}
applied to the wholesale collapse of a physical system into a
black hole of its own making. A loophole in Susskind's
argument pointed out by Wald \cite{wald} can be corrected for
quiescent systems by considering either infall of the system
into a large black hole, or a tiny auxiliary black hole
which devours the system \cite{bek00}.  The holographic bound
as above stated can be exceeded by dynamically evolving
systems, but Bousso \cite{bousso1} has given a reformulation
which works in these cases also.

The attention of the particle theory community has been
riveted on the implications of the holographic principle for
the structure of the fundamental equations of nature, for
example the equivalence between physical theories formulated
in different spacetime dimensions.  However, there is no
gainsaying the possible practical importance of fundamental
restrictions, such as the holographic one on the {\it 
information  storage  capacity\/} of systems.  For one thing,
the principles obviously serve as a final arbitrates of the
promise of any futuristic information storage technology. In
this respect the holographic bound is not an onerous one. 
For instance, it merely requires that a device with
dimensions of order a centimeter hold no more than some
$10^{66}$ bits of information.  By contrast, all the books in
the Library of Congress hold of the order of $10^{15}$ bits
of information, and no state-of-the-art one-centimeter size
memory can hold all that.  The holographic bound is thus too
liberal for the present and foreseeable technology. 
So one question to be asked is, can one device a tighter
bound on information storage than the holographic one ?  In
light of the explosive development of fast communications, a
further interesting question would be: what fundamental bounds
can be set on the {\it flow\/} of information ?  In this
lecture I shall go into such questions wherever they infringe
into the realm of gravity, which I here take to be described
by standard general relativity.

\subsection{Information and Entropy}
\label{sec:I&E}

It is plain that information can be stored in a system, by
man or by nature, only if that system has more than one state
accessible for the task.  Nothing can be learned from a
system that looks the same under all circumstances.  With two
{\it distinguishable\/} states with no bias between them, one
can already store and retrieve information, for example, the
answer to a ``yes or no'' question.  A two-state system can
thus hold a {\it bit\/} of information.  What about a $N$
state system with no bias between states ?  It makes sense to
assign it information capacity $\log_2 N$ bits.  For one
thing, this reduces to the previous case for $N=2$.  In
addition, if we have two well separated classical systems, A
and B, with $N_{\rm A}$ and
$N_{\rm B}$ (unbiased) states, respectively, there are a
total of $N_{\rm A} N_{\rm B}$ overall states, so we would
assign the joint system information capacity $\log_2(N_{\rm
A} N_{\rm B})$.  But this is just the sum of information
capacities for A and for B; thus is the logarithm function
singled out as the relevant one for quantifying information. 
The states mentioned must be precisely distinguishable.  In
classical physics they may be, for example, mutually
nonoverlapping cells in the phase space of a multiparticle
system.  In quantum theory they must be mutually orthogonal
states, because nonorthogonal states cannot be distinguished
with certainty by any measurement \cite{peres}.

What if the states are biased ?  For instance, it might be
that before measurement is made, state 1 is twice as
probable as state 2.  Does this affect the information
capacity ?  Let us imagine that our system has
$N$ equally probable states, but we divide these in groupings
of $N_1, N_2, \cdots\, $ states with $\sum_i N_i=N$.  If our
experimental resolution does not permit us to peer into the
groupings (or we just do not care to do so), we have to
consider them as regular states with probabilities
$p_1=N_1/N, p_2=N_2/N, \cdots\ $.  It is reasonable to
expect that the information capacity of the system, call it
$I_{\rm max}$,  should depend only on the set $\{p_i\}$:
$I_{\rm max}=I_{\rm max}(p_1,p_2,\cdots\,)$.  However, had we
insisted and succeeded in peering into the groupings, we
could have obtained  additional information $\log_2 N_1$ from
the first grouping (which turns up with probability $p_1$),
and so on.  Therefore, the {\it total\/} information capacity
of the system can be written in two ways which must agree:
\begin{equation} 
I_{\rm max}(p_1,p_2,\cdots\,) + \sum_i p_i\,
\log_2 N_i = \log_2 N
\end{equation} Using the normalization $\sum_i p_i =1$ it is
easy to solve for $I_{\rm max}$:
\begin{equation} I_{\rm max}(p_1,p_2,\cdots\,) = -\sum_i
p_i\, \log_2 p_i
\label{S_shannon}
\end{equation} 
This is Shannon's famous 1948 formula
\cite{shannon} for the peak information capacity of a system
with distinguishable states which occur with {\it a priori\/}
probabilities $\{p_i\}$.  It clearly makes no difference for
the final result whether the states $\{i\}$ are composite or
``elementary'', so long as the probabilities $p_i$  assigned
them reflect an operational expectation.

Shannon's $I_{\rm max}$ looks like Boltzmann's expression for
the entropy of a gas, apart from the facts that Boltzmann used
natural logarithms, and that he prefaced the expression with
the constant $k$, an historical accident stemming from the
use of different units for temperature and energy.  Beneath
these superficialities the two are the same expression:
thermodynamic entropy {\it is\/} the information storage
capacity of matter.  And thermodynamics is derivable from
information--theoretic concepts \cite{katz,peres}.  The
endless controversies in the literature around this
information--entropy equivalence seem to stem from confusion
about the level at which the probabilities
$\{p_i\}$ are formulated.  The chemist, for instance,
determines the entropy
$S$ of a piece of iron by methods that reach down to the
atomic level; for him the states $\{i\}$ are atomic states. 
The engineer, by contrast, is interested in storing
information in the magnetic domains of the iron in a magnetic
tape.  He groups atomic states into domain states which give
him new probabilities $p_i{}'$.  The $I_{\rm max}$ he
calculates from these is much smaller than the chemist's $S$,
but there is no question of conceptual contradiction.

Likewise, confusion has arisen about the relation between von
Neumann's entropy in quantum theory, $S=-{\rm Tr}\, \rho\ln
\rho$ ($\rho$ is the density matrix), and Shannon's $I_{\rm
max}$.  Apart from an obvious factor $\ln 2$, the former is
identical to $I_{\rm max}$ when the $p_i$'s in the latter
refer to the probabilities to measure values of an observable
whose eigenstates coincide with those of $\rho$ (for example
energy when
$\rho$ stands for a thermal state).  But it is possible to
design quantum measurements for which the probabilities will
give a Shannon $I_{\rm max}$ which exceeds von Neumann's $S$. 
In this case the states are no longer mutually orthogonal, and
so are not fully distinguishable, and the maximum information
that can be read out of the system is no longer equal to
(\ref{S_shannon}), but is bounded by $S\log_2 e $
\cite{peres}.  At any rate, there is {\it never\/} a
question of  $I_{\rm max}$ and $S$ being unrelated.      	

The chemist and engineer in the previous discussion are
concerned with two levels of probabilities.  In reality there
are many: matter is made of atoms which are themselves built
of electrons and nuclei.  The last are put together from
nucleons which are themselves composites of various quarks
and gluons.  There are states, and so probabilities and an
entropy, at every level of this hierarchy.   Obviously, the
deeper we go, the higher the entropy.  In what follows we
shall be interested in the entropy (information capacity)
$S_{\rm X}$ calculated at level X, the deepest level of
structure.  For ordinary matter this means the level of
lepton, quark and gluon degrees of freedom, or something
deeper if the standard model is up for a big revision. 
Obviously this $S_{\rm X}$ bounds from above the information
capacity of material media accessible with any actual
technology.

\subsection{The Entropy Bound}
\label{sec:newbound}
    
Consider in asymptotically flat
four--dimensional spacetime ($D=n+1=4$) a finite system
${\cal U}$ with energy $E$ which, for the moment, we think of
as spherical with radius $R$ ($E$ and $R$ measured in proper
frame).  How big can its entropy $S$ be ?  The holographic
principle limits it to $S\leq 4\pi R^2/\hbar$ (this is in
natural logarithm units, nits, rather than base--two
logarithm units, bits).  Can we do better without knowing
more details about ${\cal U}$ ?  Indeed, we have the entropy
bound
\begin{equation} 
S\leq  2\pi ER/\hbar
\label{original}
\end{equation}  
proposed prior to the advent of holography
\cite{bek81}.   So long as
${\cal U}$'s self-gravity is not strong, $E\ll R$.  For
example, for laboratory sized systems $E< 10^{-23} R$, while
for astronomical systems, barring neutron stars and black
holes,  $E< 10^{-5} R$.  Thus, with few exceptions, the
entropy bound is many orders of magnitude tighter than the
holographic one.  It restricts the information capacity  of
a one--centimeter device made of ordinary matter to be less
than $10^{37}$ bits, which limit no longer looks
unreachable.   

In the original derivation of bound (\ref{original}) I
imagined that
${\cal U}$ is lowered slowly from far away to the horizon of
a stationary black hole, while all the freed potential energy
is allowed to do work on a distant agent (a Geroch process
\cite{bek73}).  I then applied the GSL to get the bound
\cite{bek81}. This derivation was criticized \cite{UW} for
not taking into account the buoyancy of ${\cal U}$ in the
Unruh radiation surrounding it by virtue of its
acceleration.  A protracted controversy \cite{UW,resp} on
this issue led to the perception \cite{bek94} that correction
for buoyancy---itself an intricate calculation---merely
increases the $2\pi$ coefficient in Eq.~(\ref{original}) by a
tiny amount provided only that one assumes that
$R\geq\hbar/E$.  Now elementary particles and composite
objects do obey this restriction.  (Some solitons in $D=1+1$
dimensional theories fail to so, but solitons are not
expected to bear any entropy anyway.)  Although originally
meant only for weakly gravitating systems, bound
(\ref{original}) is actually saturated by all Kerr-Newman
black holes \cite{bek81,zaslavskii,hod,bek_mayo} provided one
interprets $E$ as the black hole's mass and $R$ as the
Boyer-Lindquist coordinate of the horizon, $r_+$.

The entropy in bound (\ref{original}) is the $S_{\rm X}$ of
Sec.~\ref{sec:I&E}, and not entropy calculated down to a
shallower level of structure.  This is because gravitation
plays a crucial role in many generic ways of deriving the
entropy bound \cite{bek81,resp,bek94,MG7}.   And gravitation
is unique among the interactions in that it is aware of all
degrees of freedom in its sources (according to the
equivalence principle all energy gravitates). How then could
bound (\ref{original}) take into account only entropy
corresponding to intermediate degrees of freedom,  and so
ignore energy--carrying states at the deeper levels ?

Only of late has the ubiquitous role of bound
(\ref{original}) and closely related ones been realized.  For
instance, Bousso \cite{bousso00} has shown, via the Geroch
process argument, that bound (\ref{original}) also applies
{\it verbatim\/} in any asymptotically flat spacetime of 
$D=n+1$ dimensions.  Further, he shows that when $E$ is
expressed in terms of ${\cal U}$'s gravitational radius
$r_{\rm g}$ given by
\begin{equation} 
r_{\rm g}{}^{n-2}={8\Gamma(n/2)
E\over(n-1)\pi^{n/2-1}\ },
 \label{rg}
\end{equation} 
the bound takes the form
\begin{equation} 
S\leq {(n-1)\pi^{n/2}\,r_{\rm g}^{n-2}R\over
4\Gamma(n/2)\,\hbar^{(n-1)/2}}. 
\label{bousso}
\end{equation}

Bousso has also established a new entropy bound,
the D--bound, for systems in $D=n+1$ de Sitter  spacetime
which occupy a small part of the space inside the
cosmological horizon (radius $r_{\rm c}$).  It is given again
by (\ref{bousso}) with $R$'s role played by $r_{\rm c}$. 
Spherical {\it black holes\/} in $D\geq 4$ spacetime (for
which the horizon has a $(n-1)$-D ``area'') also obey
(\ref{original}) in asymptotically flat spacetimes, and the
D--bound in asymptotically de Sitter spacetimes, but no
longer saturate these bounds for $D>4$ \cite{bousso00}.  

And somewhat earlier E. Verlinde \cite{verlinde} proposed
that the entropy $S$ of a complete closed Robertson--Walker
universe in $D=n+1$ spacetime dimensions whose contents are
described by a conformal field theory (CFT)---the deeper
description of a number of massless fields, possibly in
interaction---with large central charge (essentially many
particle species), is subject to the generic bound
\begin{equation} 
S\leq {2\pi R\over n\hbar }[E_{\rm
C}(2E-E_{\rm C})]^{1/2}
\label{verlinde_bound}
\end{equation} 
where $R$ is the radius of the $S^n$ space,
$E$ the total energy in the fields and $E_{\rm C}$ the
Casimir (vacuum) energy.  As Verlinde shows, this bound is a
straight generalization to $n$ space dimensions of Cardy's
famous expression for the entropy of a $D=1+1$ CFT
\cite{cardy}.  Verlinde points out that for fixed $E$ the
maximum of his bound is $2\pi R E/(n\hbar)$, which never
exceeds the original entropy bound (\ref{original}); indeed
Verlinde adopts $S \leq 2\pi R E/(n\hbar)$ as the fiducial
form of that bound.   A number of recent papers
have substantiated Verlinde's bound and so
culminate years of protracted efforts by many to make
meaningful statements about the entropy (and thus the maximum
information) that can be contained in a whole universe
\cite{more}.   

Bounds like (\ref{original}) are thus of wide applicability. 
But can one see why a bound like this should be true, at
least in some mundane context, without getting embroiled in
all the intricacies mentioned ?   Indeed, there is an easy
{\it gedanken\/} experiment which lets us do this. Drop
${\cal U}$ into a Schwarzschild black hole of mass  $M\gg E$
from a  large distance
$d\gg M$ away; $d$ is so chosen that the Hawking radiance
carries away energy (as measured at infinity) equal to $E$
while ${\cal U}$ is falling to the horizon where it is
effectively assimilated by the black hole.   At the end of the
process the black hole is back at mass $M$ and its entropy
has not changed.  Were the emission reversible, the radiated
entropy  would be
$E/T_{\rm H}$ with $T_{\em H}\equiv \hbar(8\pi M)^{-1}$. 
Curved spacetime makes the entropy emitted a factor $\nu$
larger; typical values, depending on particle species, are
\cite{page2} $\nu=1.35$--$1.64$.  Thus the overall change in
world entropy is     
\begin{equation}
\delta S_{\rm ext} = \delta S_{\rm rad}-S= \nu E/T_{\rm H} -
S  
\label{netchange} 
\end{equation} 
One can certainly  choose  $M$ larger than
$R$, say, by an order of magnitude so that the system
will fall into the hole without being torn up: $M=\zeta R$
with $\zeta = {\rm a\ few}$. Thus by the {\it ordinary\/}
second law we obtain the bound
\begin{equation}  
S < 8\pi\nu\zeta RE/\hbar   
\label{newbound}   
\end{equation} 
This bound applies to a rather arbitrary
system ${\cal U}$ in terms of its total energy $E$ and size
$R$.   But in the derivation ${\cal U}$ is not allowed to be
be strongly gravitating (meaning $R\sim E$) because then $M$
could not be large compared to $E$, as we have assumed, while
$\zeta$ is of order a few.  We thus have to assume in
addition $R\gg E$.

Note that we could not derive (\ref{newbound}) by using a
heat reservoir in lieu of a black hole.  A reservoir which
has gained energy $E$ upon ${\cal U}$'s assimilation, and has
returned to its initial energy by radiating, does not
necessarily return to its initial entropy, certainly not
until ${\cal U}$ equilibrates with the rest of the
reservoir.  But a (nonrotating uncharged) black hole whose
mass has not changed overall, retains its original entropy
because that depends only on mass.  In addition, for the
black hole mass and radius are related in a simple way; this
allowed us to replace $T_{\rm H}$ in terms of $R$. By
contrast, for a generic reservoir, size is not simply related
to temperature.

Of course, the above derivation ignores the effect of Hawking
radiation pressure.  How important is this ?  Could it blow
${\cal U}$ outwards ?   If we approximate the radiance as
black--body radiance of temperature $T_{\rm H}$ coming from a
sphere of radius $2M$, the energy flux at Schwarzschild
coordinate
$r$ from the hole is 
\begin{equation} 
F(r)= {{\cal N}\hbar\over 61,440(\pi M r)^2},
\label{flux}
\end{equation} 
where ${\cal N}$ stands for the effective
number of massless species radiated (photons contribute 1 to
${\cal N}$ and each neutrino species $7/16$).  This estimate
is known to be off by a factor of only a few \cite{page1}. 
This energy (and momentum) flux results in a radiation
pressure force $f_{\rm rad}(r)=\pi R^2 F(r)$ on ${\cal U}$. 
More precisely, species which reflect well off ${\cal U}$ are
approximately twice as effective at exerting force as just
stated, while those (neutrinos and gravitons) which go right
through
${\cal U}$ contribute very little; the ${\cal N}$ must thus
be reinterpreted accordingly.  I have ignored relativistic
corrections so that the result, as qualified, is correct
mostly for $r\gg M$.  

Writing the gravitational force on ${\cal U}$ in the 
Newtonian approximation,  $f_{\rm grav}(r)=ME/r^2$, one sees
that
\begin{equation} 
{f_{\rm rad}(r)\over f_{\rm grav}(r)}=
{{\cal N}_{\rm eff}\,\hbar R^2\over 61,440  \pi^2 M^3 E}
\label{ratio}
\end{equation} 
I have written ${\cal N}_{\rm eff}$ here
because, as mentioned, some species just pass through ${\cal
U}$ without exerting force on it.  In addition, only those
species actually represented in the radiation flowing out
during ${\cal U}$'s infall have a chance to exert forces. 
Now an Hawking quantum bears an energy of order $T_{\rm H}$,
so the number of quanta radiated together with energy $E$ is
approximately  $8\pi M E/\hbar$.  By our assumption that 
$\hbar/E < R$ and our stipulation that
$M=\zeta R>R$, this is large compared to unity.  Since a
species can exert pressure only if it is represented by at
least one quantum, one obviously has
${\cal N}_{\rm eff} < 8\pi ME/\hbar$.  Therefore,
\begin{equation}   
{f_{\rm rad}(r)\over f_{\rm grav}(r)}<
{R^2\over 7680  \pi M^2}
\ll 1 
\label{newratio}
\end{equation} 
Radiation pressure is thus negligible, and
${\cal U}$'s fall is very nearly on a geodesic, at least
until ${\cal U}$ approaches to within a few Schwarzschild
radii.  It is intuitively clear that if $d\gg M$, the last
(relativistic) stage cannot make any difference, and ${\cal
U}$ must plunge to the horizon.

Whether $d$ is large enough must be checked.  We have taken
it such that the infall time equals the time $t$ for the hole
to radiate energy $E$.  Newtonially  $d\approx 2(t^2
M/\pi^2)^{1/3}$, while Eq.~(\ref{flux}) gives the estimate
$t\approx 5\times 10^4 EM^2\hbar^{-1}{\cal N}^{-1}$ with
${\cal N}$ now the full species number.  From these equations
and $M=\zeta R$ we get that $d\approx 1.2\times 10^3 (\zeta
ER/{\cal N}\hbar)^{2/3}M$.  Thus for ${\cal N}<10^2$
(conservative estimate of {\it our\/} world's massless
particle content), we have $d>57 M$ for all systems ${\cal
U}$ satisfying our assumption $R>\hbar/E$.  Thus for all
these we have established the entropy bound
(\ref{newbound}).  

Our simple argument here leaves the factor $\nu\zeta$
somewhat fuzzy; but it is safe to say that $4\nu\zeta<10^2$. 
Thus we recover bound (\ref{original}) with some overshoot of
the coefficient, not a large prize to pay for the simplicity
of the derivation.   When we come to strongly gravitating
systems ($E\sim R$), we cannot derive the bound
(\ref{original}) or even the weaker version (\ref{newbound})
by the methods just expounded. Nevertheless, as mentioned, a
black hole in $D=4$ spacetime saturates bound
(\ref{original}) and complies with it for $D>4$.   For
strongly gravitating systems in asymptotically flat
spacetime, the holographic bound and the entropy bound make
very similar predictions  for $D=4$, but for $D>4$ the
holographic bound is the tighter of the two.  Unless $D$ is
very large,  the entropy bound is the tighter bound for
weakly gravitating systems, such as those we meet in everyday
life.

\subsection{Black Holes as One--Dimensional Information
Conduits}
\label{sec:transfer}

The holographic bound (\ref{holographic}) is supposed to be
telling us that a generic physical system in 4--D spacetime
is fundamentally two--dimensional in space.   It turns out
that viewed as an information absorber or entropy emitter, a
black hole in 4--D spacetime is fundamentally
one--dimensional in space \cite{bek_mayo01}.  

To show this one must define a one--dimensional information
transmitting system---a channel.  In flat spacetime a
channel is a complete set of one-way propagating modes of
some field, with the modes enumerated by a single parameter. 
For example, all electromagnetic modes in free space with
fixed wave vector {\it direction\/} and particular linear
polarization constitute a channel; the modes are parametrized
solely by frequency.  One might implement such a channel with
a straight infinitely long coaxial cable (which is well known
to transmit all frequencies) capped  at its entrance by the
analog of a polaroid filter.  Acoustic and neutrino channels
can also be defined.  A fundamental question is: what is the
maximum {\it rate\/}, in quantum theory, at which information
may be transmitted down such a channel for prescribed power
$P$ ?   The answer was found in the 1960's by several
information theory pioneers (see my review with
Schiffer\cite{bek_schiff}),
but I want to reproduce here the much later but very simple
derivation of Pendry \cite{pendry}, which is of very broad
applicability.

Pendry thinks of a possible signal state as corresponding to
a particular set of occupation numbers for the various 
propagating modes.  He assumes the channel is uniform in the
direction of propagation, which allows him to label the modes
by momentum $p$.  But he allows for dispersion, so that a
quantum with momentum $p$  has some energy $\varepsilon(p)$.
Then the propagation velocity of the quanta is the group
velocity $\upsilon(p)=d\varepsilon(p)/dp$.  Up to a factor 
$\ln 2$ the information rate capacity is just the maximal
one-way entropy current for given
$P$, which obviously occurs for the thermal state, if one
excludes the modes moving opposite the direction of interest. 

Now the entropy $s(p) $ of any boson mode of momentum $p$ in
a thermal state (temperature $T$) is \cite{LL}
\begin{equation} 
s(p)={\varepsilon(p)/T\over
e^{\varepsilon(p)/T}-1}-\ln \left(
1-e^{-\varepsilon(p)/T}\right),
\label{s}
\end{equation} 
so the entropy current in one direction is
\begin{equation}
\dot S=\int^{\infty}_0 s(p)\thinspace\upsilon(p)\, {dp\over
2\pi\hbar}= \int^{\infty}_0 s(p)\thinspace {d\varepsilon\over
dp}\ {dp\over 2\pi\hbar},
\label{current}
\end{equation} 
where $dp/2 \pi \hbar $ is the number of modes
per unit length in the interval $dp$ which propagate in one
direction.  This factor, when multiplied by the group
velocity, gives the one-way current of modes.   Suppose
$\varepsilon(p)$ is monotonic and extends over the range
$[0,\infty)$; we may then cancel $dp$
and integrate over $\varepsilon$.  Then after substitution of
Eq.~(\ref{s}) and integration by parts we have
\begin{equation}
\dot S= {2\over T}\int^{\infty}_0 {\varepsilon\thinspace \over
e^{\varepsilon/T} -1}\ { d\varepsilon\over 2\pi\hbar} ={2\over
T}\int^{\infty}_0 {\varepsilon(p)\over e^{\varepsilon(p)/T}
-1}\thinspace
\upsilon(p) \thinspace {dp\over 2\pi\hbar }. 
\label{entropy_flow}
\end{equation} 
The first factor in each integrand is the mean
energy per mode, so that the integral represents the  one-way
power $P$ in the channel.  Thus 
\begin{equation}
\dot S=2P/T . 
\label{final}
\end{equation}

The integral for $P$ in the first form of
Eq.~(\ref{entropy_flow}) can easily be done:
\begin{equation} 
P = {\pi (T)^2\over 12 \hbar}.
\label{power}
\end{equation} 
Eliminating $T$ between the last two
expressions gives Pendry's limit
\begin{equation}
\dot S=\left({\pi P\over 3\hbar}\right)^{1/2}\qquad {\rm
or}\qquad \dot I_{\rm max}= \left({\pi P\over
3\hbar}\right)^{1/2}\log_2 e.
\label{pendry_formula}
\end{equation} 
For a fermion channel $P$ in Eq.~(\ref{power})
is a factor 2 smaller, and consequently $\dot S$ in
Eq.~(\ref{pendry_formula}) is reduced by a factor
$\surd 2$.
 
The function $\dot S(P)$ in Eq.~(\ref{pendry_formula}) is
the so called {\it noiseless quantum channel capacity\/}. 
Surprisingly it is independent, not only of the form of the
mode velocity $\upsilon(p)$, but also of its scale.  Thus the
phonon channel capacity is as large as the photon channel
capacity despite the difference in speeds. Why? Although
phonons convey information at lower speed, the energy of a
phonon is proportionately smaller than that of a photon in
the equivalent mode.  Thus when the capacities of channels
harnessing various carriers are expressed in terms of power,
they turns out to involve the same constants.  
Formula.~(\ref{pendry_formula}) neatly characterizes what we
mean by one--dimensional transmission of entropy or
information.  It refers to transmission by use of a single
species of quantum and a specific polarization; different
species and alternative polarizations engender separate
channels.  Although framed in a flat spacetime context, its
lack of sensitivity to the dispersion relation of the
transmitting {\it milieu\/} should make Pendry's limit
relevant to curved spacetime also.  This because
electrodynamics in curved spacetime is equivalent to flat
spacetime electrodynamics in a suitable dielectric and
paramagnetic medium \cite{volkov}. 

By contrast the power and entropy emission rate in a single
boson polarization of a closed black body surface with 
temperature $T$ and area $A$  in flat 4--D
spacetime is (half the Stefan--Boltzmann law) 
\begin{equation} 
P={\pi^2 T^4 A\over 120 \hbar^3}\qquad\quad \dot S ={4\over
3}{P\over T}
\label{P}
\end{equation} 
 whereby
\begin{equation}
\dot S= {2\over 3}\left({2\pi^2AP^3\over
15\hbar^3}\right)^{1/4}
\label{3-D}
\end{equation} 
[for fermions $P$ carries an extra factor
$7/8$ and formula (\ref{3-D}) an extra factor
$(8/7)^{1/4}$].   Our manifestly 3--D transmission system
deviates from the sleek formula (\ref{pendry_formula}) in the
exponent of $P$ and in the appearance of the measure $A$ of
the system.  In emission from a closed curve of length $L$ in
two--dimensional space, the factor $ (L P^2)^{1/3}$ would
replace $(A P^3)^{1/4}$.  We may thus gather the
dimensionality of the transmission system from the exponent
of $P$ in the expression $\dot S(P)$ [it is  $n/(n+1)$ for
$D=n+1$ spacetime dimensions], as well as from the value of
the coefficient of $P/T$ in expressions for $\dot S$ like
(\ref{final}) or (\ref{P}) [it is
$(n+1)/n$].

Radiation from a Schwarzschild black hole in 4--D spacetime
is also given by Eqs.~(\ref{P}) (or their
fermion version) with $A=4\pi (2M)^2$ and $T=T_{\rm H}$,
except we must correct the expression for
$P$ by a species dependent factor $\bar\Gamma$ of order unity
\cite{page1}, and replace the $4/3$ in the expression for 
$\dot S$ by the species dependent factor $\nu$ already
mentioned in Sec.~\ref{sec:newbound}.  Eliminating $M$
between the equations we obtain, in lieu of Eq.~(\ref{3-D}),
\begin{equation}
\dot S = \left({\nu^2\bar\Gamma\pi P\over
480\hbar}\right)^{1/2}.
\label{BHlimit}
\end{equation} 
(For fermions there is an extra factor $7/8$
inside the radical). This looks completely different from the
law (\ref{3-D}) for the hot closed surface because, unlike
for a hot body, a black hole's temperature is related to its
mass.

However, (\ref{BHlimit}) {\it is\/} of the same form as
Pendry's limit (\ref{pendry_formula}) for one--channel
transmission.  From Page \cite{page1,page2}
we get $\bar\Gamma=1.6267$ and  $\nu=1.5003$ for one photon
polarization, so the numerical coefficient of (\ref{BHlimit})
is $15.1$ times that in (\ref{pendry_formula}).  Repeating the
above exercise for one species of neutrinos we again find
formulae like (\ref{BHlimit}) and  (\ref{pendry_formula}),
this time with
$\bar\Gamma=18.045$ and  $\nu=1.6391$;  the  numerical
coefficient of (\ref{BHlimit}) is $48.1$ times that of the
fermion version of (\ref{pendry_formula}). 

Evidently in its entropy emission properties a black hole in
4--D spacetime is more like a 1--D channel than like a
surface in 3--D space. Why is this ?  A formal answer is
that, because of the way $T_{\rm H}$ is related to the black
hole's radius $2M$,  Hawking emission prefers to emerge in
the lowest angular momentum mode possible.  To exit with
angular momentum $j\hbar$, a quantum must have energy
(momentum) $\hbar\omega$ of order
$j\hbar/2M$.  But in the Hawking thermal distribution the
dominant $\hbar\omega$ is of order  $T_{\rm H}=\hbar(8\pi
M)^{-1}$.  Thus the emerging $j$'s tend to be small.  For
example, $97.9\%$ of the photon energy emerges in the $j=1$
modes ($j=0$ is forbidden for photons), and $96.3\%$ of the
neutrino power is in the
$j={\scriptstyle 1\over\scriptstyle 2}$ modes \cite{page1}. 
Thus the black hole emits as close to
radially as possible.  This means that, crudely
speaking, it does so through just one channel.

\subsection{Information Pulses in Curved Spacetime}
\label{sec:bursts}

The discussion in Sec.~\ref{sec:transfer} centered on steady
state streams of information and energy.  What if information
is delivered as pulses ?  Can one state a bound generalizing
(\ref{pendry_formula}) ?  Further, can one include effects of
gravitation on information transfer rate ?  To answer these
questions let us extend the notion of channel to curved
spacetime, at least to stationary curved spacetime.  Again, a
channel will be a complete set of one-way modes of some field
that can be enumerated with a single parameter.  Each channel
is characterized by species of quanta, polarization
(helicity), trajectory, etc.  In Sec.~\ref{sec:transfer} we
characterized the {\it signal\/} in a particular channel by
power.  For a pulse it seems a better idea to use as signal
parameters the signal's duration $\tau$ and its energy $E$. 
Since in curved spacetime a channel is not generally uniform,
we choose to measure these parameters in a local Lorentz frame
(we shall see presently that it does not matter which Lorentz
frame).  This precaution allows us to focus on sections of
the channel and treat them as if we were working in flat
spacetime.

How is $ I_{\rm max}$ related to $E$ and $\tau$ ?  Since
information is dimensionless, $I_{\rm max}$ must be a function
of dimensionless combinations of $E$,
$\tau$, channel parameters and fundamental constants. We
exclude channels which transmit massive quanta, e.g.
electrons, because rest mass is energy in a form not useful
for communication, so  that the strictest limits on
$I_{\rm max}$ should emerge for massless signal carriers. 
Hence Compton lengths do not enter into the argument. Also in
order to maximize the information flux, we focus on broadband
channels, and exclude any frequency cutoff.  Finally, we
exclude situations where the signal undergoes dispersion;
this has the practical upshot that apart from light's speed
$c$, only one other velocity---the signal velocity
$c_s$---can appear.  We consider $c_s/c$ a property of the
{\it channel\/} because it is common to all
signals.  If we {\it temporarily\/} exclude the gravitational
constant, there is thus a single dimensionless combination of
{\it signal} parameters,
$\xi=E\tau/\hbar$, at our disposal. Thus
\begin{equation} 
I_{\rm max}=\Im (E\tau/\hbar), 
\label{imax}
\end{equation} 
where $\Im(\xi)$ is some nonnegative valued
function characteristic of the  channel, the characteristic
information function (CIF) \cite{bek_schiff}.

Let us check formula (\ref{imax}) in flat spacetime in 
steady state (momentarily return to a
long stream of information).  Steady state means that the
signal can be characterized as statistically stationary in a
suitable frame.  It should thus be possible to infer the peak
information transfer {\it rate\/} by focusing on a finite
section of the signal bearing information  $I_{\rm max}$ and
energy $E$. It should matter little how long a stretch in
$\tau$ is  used so long as it is not too short. This can only
be true if $\dot I_{\rm max}\equiv E\tau^{-1}$ is fully
determined by  the power $P\equiv E\tau^{-1}$, and this is
consistent with Eq.~(\ref{imax}) only if
$\Im(\xi)\propto\surd \xi$ for only then does $\tau$ cancel
out. We have thus recovered Pendry's limit
(\ref{pendry_formula}), the correct answer
for steady state; formula (\ref{imax}) checks out.  

The dividing line between steady state information transfer
and transfer by  means of very long signals is not sharp. This
suggests that long pulse signals must also obey a Pendry type
formula, albeit approximately
\cite{marko}. The law  $\dot I_{\rm max}\propto
(P/\hbar)^{1/2}$ is evidently inapplicable to {\it brief\/}
information pulses.  For such it may be replaced by a linear
upper bound
\cite{bek81b} which may even transcend some of the
limitations we imposed to define $\Im(\xi)$.  Consider the
information $I$ to be encoded in some material structure
${\cal V}$ of radius $R$ and rest energy $E$ which maintains
its integrity and dimensions as it flies from emitter to
receiver.  From Eq.~(\ref{original}) we have
the strict inequality $I<2\pi E R
\hbar^{-1}\log_2 e$.  The rate at which the information is
assimilated by the receiver is obviously restricted by the
local time
$\tau$ it takes for ${\cal V}$ to sweep by it.  From special
relativity
$\tau> 2R\gamma^{-1}$ with $\gamma$ accounting for the
Fitzgerald contraction of ${\cal V}$ in the frame of the
receiver.  Thus the peak information reception rate is
$I/\tau< \pi E \hbar^{-1}\log_2 e$, or
\begin{equation}
\dot I_{\rm rec}< \pi E_{\rm rec}\hbar^{-1}\log_2 e
\label{linear}
\end{equation} 
where $E_{\rm rec}\equiv \gamma E$ is ${\cal
V}$'s energy as measured in the receiver's frame.  This
replaces the information version of
Eq.~(\ref{pendry_formula}) when it comes to pulses.   With
$\xi\equiv E_{\rm rec}\tau\hbar^{-1}$ we thus have the strict
linear bound $\Im(\xi)< (\pi \log_2 e)\xi $.  There is a lot
of  evidence \cite{bek88,bek_schiff} that this bound
applies even when the signal has no rest frame.  One should
not be alarmed because the law
$\Im(\xi)\propto \surd\xi\ $ figuring in
(\ref{pendry_formula}) exceeds the linear bound
for {\it small\/} $\xi$; the law is meant for
steady state, which makes sense only in the limit
$\xi=E\tau/\hbar\rightarrow
\infty$. 

Let us check the {\it local\/} Lorentz invariance
of (\ref{imax}).  Consider a pair of local
Lorentz frames, A and B, encompassing a section of the
channel, with B moving to the right with respect to A  with
speed $V<c_s$, and let $\gamma=(1-V^2)^{1/2}$.  {\it If\/}
there is a medium, A is taken as its rest frame, and $c_s$ is
the signal's speed in this frame.  Now let a right moving
signal pulse's front (speed
$c_s)$ pass the origins of these frames at time
$t_{\rm A}=0$ when they coincide.  At a later
time
$t_{\rm A}=t_1$ the pulse's rear has reached A's origin; by
then B's origin is at $x_{\rm A}=Vt_1$.  Sometime later, at
$t_{\rm A}=t_2$, the pulse's rear reaches B's origin which is
then at $x_{\rm A}=Vt_2$.  Calculating the pulse's length in
frame A in two ways gives
$(c_s-V)t_2=c_s t_1$, so that
\begin{equation} t_1/t_2=1-V/c_s
\label{f}
\end{equation} 
The signal duration is $\tau_{\rm A}=t_1$ in
A; in B, however, it is
$\tau_{\rm B}=t_2\gamma^{-1}$ because $t_2$ is the time it
takes in A for all the pulse to traverse B's origin; thus
\begin{equation}
\tau_{\rm A}=\gamma\tau_{\rm B}(1-V/c_s).
\label{g}
\end{equation}

Now in A the energy and momentum of a (massless) quantum in
the signal stand in the ratio $c_s$.  Thus if interactions
are negligible, the energy  $E_{\rm A}$ and the corresponding
momentum $P_{\rm A}$ of the full pulse stand also in the same
ratio.  Considering a Lorentz boost to frame B we discover
that
\begin{equation} 
E_{\rm B}=\gamma (E_{\rm A}-VP_{\rm
A})=\gamma E_{\rm A}(1-V/c_s).
\label{after}
\end{equation} 
Therefore, $E_{\rm B}\tau_{\rm B}= E_{\rm
A}\tau_{\rm A}$.  We can likewise prove this result for $1>V>
c_s$.  In view of this and the fact that information is a
Lorentz scalar, the statement (\ref{imax}) is seen to be
Lorentz invariant.  This has immediate applications.  For
instance, A can be interpreted as the signal receiver's frame
and B as the propagating medium's, or perhaps the
transmitter's.

But how is the limit on information transmission rate related
at two point along the channel ?   In flat spacetime, and in
the absence of dispersion, $E$ and $\tau$ are evidently
conserved with propagation.  So is the
information, so that Eq.~(\ref{imax}) is valid at every point
along the channel.  Once we are in stationary {\it curved\/}
spacetime, $E$ and $\tau$ are subject to redshift and
dilation effects, respectively.  However, the effects act in
opposite senses on $E$ and $\tau$, and since they depend on
the same metric component, $E\tau$ is again
conserved throughout the signal's flight.  Therefore,
Eq.~(\ref{imax}) is meaningful throughout the channel.  In
fact one can use global values (as measured at infinity)  of
$E$ and $\tau$ in Eq.~(\ref{imax}).    We see that {\it one
and the same  formula limits information transmission,
propagation and reception rates\/}.

We have so far excluded appearance of the gravitational
constant in the formulae.  This means we have been neglecting
self--gravity of the signal, because this is measured by the
parameter (momentarily restoring $G$ and $c$)
$\varpi\equiv GEc^{-5}\tau^{-1}$, which is on the order of
the ratio of the signal's self--potential energy to $E$, or
that of the signal's gravitational radius to its duration. 
Should we include $\varpi$ as a new argument in
Eq.~(\ref{imax}) ?  One reason for not doing so is that it
would obviously spoil the local Lorentz invariance of
Eq.~(\ref{imax}) because $E/\tau$ is not a Lorentz scalar.  We
cannot allow this for signals propagating in vacuum in a flat
background, for this would be tantamount to a violation of
special relativity.  However, it could be argued that the
presence of $\varpi$ in
$\Im$ is permissible for signals propagating in a medium
($c_s\neq c$) because it locally defines a preferred Lorentz
frame 

In a curved background there are further arguments against
inclusion of $\varpi$ in $\Im$.  In vacuum we
can use the requirement of local Lorentz invariance to bar
$\varpi$'s appearance, for a sufficiently brief signal should
admit being encompassed in its entirety by local Lorentz
frames.  Further, $\varpi$ evidently decreases as the signal
propagates {\it outward\/} in the gravitational potential. 
Thus, $\Im(E\tau/\hbar,\varpi)$ would decrease either
outwardly (if $\Im$ increases with $\varpi$) or inwardly (if
it decreases as $\varpi$ increases).  If a signal's
information saturates the bound $\Im(E\tau/\hbar,\varpi)$ at
some point in the potential, then by conservation of
information it will exceed the bound once it has propagated
somewhat in the direction in which $\Im$ decreases.  This
leads to a contradiction. One could try to resolve the
problem by defining $I_{\rm max}$ only in terms of the {\it
minimum\/} value of
$\varpi$ in the channel. But it seems strange that, at least
for brief signals, one cannot state $I_{\rm nax}$ in terms
of local quantities.  

It thus seems that for signals propagating in vacuum in
curved spacetime, $\varpi$ cannot appear in $\Im$.  It
is unclear whether this conclusion extends to signal
propagation in a medium, because in curved spacetime a medium
is never homogeneous, which means, among other things, that
$c_s$ varies.  This in itself puts in doubt our argument for
simplicity of the expression for
$I_{\rm max}$.

\subsection{Dumping Information into a Black Hole}
\label{sec:dump}
 
Suppose we have at our disposal a certain power ${\cal P}$ to
accomplish the task of getting rid of a stream of possibly
compromising information by dumping it into a black hole.  We
may pick the size of black hole which suits us best.  By
the complementary relation between entropy and information,
formula (\ref{BHlimit}) for entropy rate
{\it out\/} of the black hole suggests a bound $\dot I_{\rm
max}\sim ({\cal P}/\hbar)^{1/2}$ on the
dumping rate.  Here we verify this new
bound.   

First we argue that if the signal originates from afar, it is
transmitted more or less through a single channel (per field
species and polarization). Let us recall the rule for field
mode counting.  In one space dimension a length $L$ contains
$(2\pi)^{-1}L\Delta k$ modes in the wave vector interval
$\Delta k$.  In 3--D space we would have $(2\pi)^{-3}L_x L_y
L_z\Delta k_x \Delta k_y \Delta k_z$ modes.  From this we may
conclude that if a flat 2--surface of area ${\cal A}$
radiates into a narrow solid angle $\Delta\Omega$  about its
normal, the number of modes out to a distance $L$ from it
whose wave vector magnitudes lie between
$k$ and $k+\Delta k$ is $(2\pi)^{-3}{\cal A}L k^2 \Delta\Omega
\Delta k$.  The factor $(2\pi)^{-1}L\Delta k$ is obviously
the number of modes emitted sequentially in each direction
and distinguished by their values of $k$. One can thus think
of ${\cal W}=(2\pi)^{-2}{\cal A} k^2 \Delta\Omega$ as the
number of active {\it channels\/}. 

Now let a transmitter with effective area ${\cal A}$ send an
information bearing signal towards a Schwarzschild black hole
of mass
$M$ surrounded by vacuum and situated at distance $d\gg 2M$. 
Let
${\cal A}$ be oriented with its normal towards the black
hole.  Evidently, as viewed from the transmitter the black
hole subtends solid angle
$\Delta\Omega=\pi(2M)^2/d^2$, and we must have ${\cal A}<4\pi
d^2$.  What should we take for $k$ in the formula for ${\cal
W}$ ?  Being interested in the highest information for given
energy (other things being equal), we certainly want to use
the smallest $k$ (smallest
$\hbar\omega$) possible.  But signals composed of too small
$k$'s will just be scattered by the black hole.  The
borderline is $k=2\pi/\lambda \approx 2\pi/(2M)$. With this
we find ${\cal W}< 4\pi^2$,  which means that information
transmission down a black hole is necessarily a few channels'
affair (for each field species and polarization, of course)
regardless of the scales
$M$ and $d$ in the problem.  

In light of this we employ formula (\ref{imax}).  Further,
since $E\tau$ is conserved in Schwarzschild (stationary)
spacetime, and closely equals ${\cal E} t$, the values being
measured at infinity, we have $I_{\rm max}=\Im({\cal E}
t/\hbar)$.  This for a pulse of duration $t$ as seen from
infinity.  If we are dealing with a steady state stream of
energy and information ($t\rightarrow\infty$ and ${\cal
E}\rightarrow\infty$ with ${\cal P}\equiv
\lim ({\cal E}/t)$ finite), we have by the logic of the
paragraph following Eq.~(\ref{imax}) that the maximum
information disposal rate into the black hole is $\dot I_{\rm
max}\sim ({\cal P}/\hbar)^{1/2}$.  We have thus checked our
guess. The precise proportionality factors for various
fields may be worked out from Eq.~(\ref{BHlimit}).   At any
rate, we uncover a ``cost'' for information disposal into a
black hole: the power required grows {\it quadratically\/}
with the information dumping rate.    

\section{Black Hole Spectroscopy}
\label{BHS}

\subsection{Introduction}
\label{intro2}

\label{I1} In classical general relativity we think of a
black hole's parameters, typically charge $Q$, angular
momentum ${\bf J}$ and mass $M$ (but others are possible: see
Erick Weinberg's lectures in this collection) as continuous.
In reality its angular momentum and charge should have
discrete spectra.   And what about $M$ ?  Is it continuous as
in the classical theory, or discrete as for an atom ? 
Granted that this is a fair question only for quantum
gravity, it is still interesting to find out if something can
be said about it in the fragmentary state of knowledge today.

I have long argued \cite{bek74} that certain features of
classical black hole physics hint at a discrete mass
spectrum of a very definite form.  That would make black
holes quite like atoms in one sense.  Here I would like to
summarize the various steps in the logic towards this
conclusion, and describe an algebraic quantum formalism
\cite{brazil96,MG8,brazil98} designed to deal with this
question irrespective of how the final quantum gravity may
turn out.  I shall use geometric units with $G=c=1$; $\surd
\hbar$ denotes the Planck mass.  

\subsection{Adiabatic Invariance and Black Hole Quantization}
\label{AI}

The present situation in quantum gravity's development, with
rival theories (string theory, loop quantum gravity,
canonical quantum gravity in several versions, \dots) whose
elements can still not be set in one to one correspondence is
somewhat reminiscent of quantum physics before Schrodinger's
equation and Heisenberg's matrix mechanics.  People then were
very much guided by the rules of the old quantum mechanics. 
And an important unifying principle of these was Ehrenfest's
(Born's excellent book \cite{born} is a convenient source): a
quantity which classically is an adiabatic invariant has at
the quantum level a discrete spectrum.
The Bohr--Sommerfeld quantization rules, which focused on
Jacobi actions $\oint p\, dq$ (all of which are adiabatic
invariants), and which gave the correct hydrogen
spectrum---fine structure included---exemplify the success of
Ehrenfest's principle.

In the 1970's the work of Penrose and Floyd \cite{penrose},
Christodoulou \cite{christodoulou} and Hawking
\cite{hawking_area} taught us that transformations of a
classical black hole are generically irreversible because the
surface area of the event horizon increases in most such
changes.  While the thermodynamic ramifications of this
discovery have been of wide interest to the gravitation and
particle communities for three decades, a less visible aspect
of this purely mechanical behavior will be the center of
attention here.  

Christodoulou and Ruffini \cite{christodoulou,cr} exhibited a
set of black hole transformations which do not involve
horizon area increase.   The simplest example I can find
deals with a point charge
$\varepsilon$ which is deposited at rest, by whatever means,
alongside the horizon of a Reissner--Nordstr\"om (RN) black
hole with charge $Q$ of the same sign as $\varepsilon$.  This
means, purely on mechanical grounds, that the energy of the
particle as measured at infinity, $E$, equals
$\varepsilon\Phi$ with 
\begin{equation}
\Phi={Q\over M+ (M^2-Q^2)^{1/2}}
\label{Phi}
\end{equation} 
the electrostatic potential at the horizon. 
Now differentiation of the expression for horizon area of our
Reissner--Nordstr\"om black hole, $A=4\pi (M+
(M^2-Q^2)^{1/2})^2$, gives
\begin{eqnarray}
\delta A &=& (\delta M-\Phi\delta Q)\Theta^{-1},
\label{dA}
\\
\Theta &\equiv& {\scriptstyle 1\over \scriptstyle 2}
(M^2-Q^2)^{1/2} A^{-1}.
\label{Theta}
\end{eqnarray}  By energy and charge conservation, when the
particle is captured by the black hole, $\delta M=E$ while
$\delta Q=\varepsilon$.  But we selected
$E=\varepsilon\Phi$, so that $\delta A=0$ upon capture of the
particle. The process of charge assimilation is a slow one
since the charge is gently placed on the horizon; and it
results in a change of $M$ and $Q$ but with no consequent
change in $A$.  Hence $A$ is an adiabatic invariant here. 
Many other examples have been collected
\cite{bekbook,mayo,duez} involving the adiabatic addition of
angular momentum to the  black hole instead of charge, or
various wave perturbations.  In all these $A$ is unchanged.  

The proposed adiabatic principle is a classical one.  But if
we take into account that by quantum theory the particle to
be assimilated has an effective minimal radius (Compton
length) which prevents us from placing its center exactly at
the horizon without already loosing it, the same sort of
calculation \cite{bek73,brazil98} shows that the minimal
horizon area increase is $(\delta A)_{\rm min}=\xi\hbar$, with
$\xi$ a constant of $O(1)$.  Adiabatic invariance is thus not
literally true, but for super--Planckian black holes the area
increase is relatively small:
\begin{equation}
\left({\delta A\over A}\right)_{\rm min} = {\xi \hbar\over A}
< { 137\xi e^2\over 4\pi Q^2} < { 137\xi\over
4\pi}\left({\delta Q\over Q}\right)^2.
\label{ineq}
\end{equation} 
Here we have used the fact that the
elementary charge is of order $\surd 137\
\hbar^{1/2}$, that in all RN black holes $Q\leq
(A/4\pi)^{1/2}$, and that the change in
$Q$ cannot be less than $e$.  Inequality (\ref{ineq}) assures
us that the fractional change in horizon area is small
compared to that of black hole charge (pressumably $|\delta
Q|\ll Q$), and generally also that of black hole mass because
\begin{equation} 
{\delta M\over M}= {\Phi \delta Q\over M}
>\left({Q^2\over 2M^2}\right){\delta Q\over Q}.
\label{dM}
\end{equation} 
Thus in order for $(\delta A)_{\rm min}/A$
{\it not\/} to be small compared to $\delta M/M$, the black
hole would have to be virtually neutral.  In this sense we
still have adiabatic invariance while allowing for quantum
mechanics of matter.  In fact, adiabatic invariance of $A$
seems to survive into the quantum gravity regime
\cite{bar_das_kunst}.

From the adiabatic invariance of the area of the horizon and
Ehrenfest's principle, we should suspect that in quantum
gravity the horizon area is replaced by an operator with a
discrete spectrum.  In the RN example the classical formula
for $A$ gives $M=(A/16\pi)^{1/2}[1+4\pi Q^2/A]$.  If this
were true for operators, the discrete spectra of $Q$ and
$A$ would imply that $M$ too has a discrete spectrum
\cite{MG8}.  Of course, quantum corrections to this last
formula might come about \cite{gour} without eliminating the
discreteness of $M$.  And if the area eigenvalues are
themselves degenerate, quantum corrections might split this
degeneracy
\cite{MG8,brazil98}.  However, it seems most constructive, in
view of the absence of a consensus quantum gravity theory, to
consider the pristine situation before all these corrections
are effective. 

\subsection{Dynamical variables and creation operators}
\label{Dyn}

In quantum theory ${\bf J}$ and $Q$ are represented by
hermitian operators $\hat {\bf J}$ and $\hat Q$.  We assume
$[{\bf
\hat J}, \hat Q] = 0$ so that the black hole can
simultaneously have sharp charge, which we assume to be an
integer multiple of $e$, and angular momentum.  In order to
get usual  spectrum  ${\bf J}^2=j(j+1)\hbar^2;\ j=0,
{\scriptstyle1\over \scriptstyle 2}\hbar, \hbar \cdots$ with
$J_z\equiv m\hbar =\{-j, -j+1, \cdots, j\}\hbar$, we have to
assume that ${\bf \hat J}\times {\bf \hat J}=\imath\hbar {\bf
\hat J}$.  The argument at the end of Sec.~\ref{AI}
predisposes us to expect that there is some observable
representing horizon area, $\hat A$, which has a simple
spectrum, $\{ a_1, a_2,  \cdots a_n\cdots \,\}$ with
$a_{n+1}>a_n$.   Now the horizon area of a black hole is
invariant under rotations of its spin; since ${\bf \hat J}$
is the generator of such rotations in quantum theory, one
expects that $[\hat A,
\hat{\bf J}] = 0$.  Similarly, horizon area is invariant
under gauge transformations; in quantum theory their
generator is, as usual, the charge
$\hat Q$.  Hence we expect that  $[\hat A, \hat Q] = 0$.   

It follows that we can conceive of a basis of one-black hole
states of the form  $\{|njmqs\rangle\}$ where $q$ is an
integer eigenvalue of
$\hat Q/e$, and $s$ distinguishes between the different black
hole states with like  $a_n$, $j, m$ and $q$ (this degeneracy
is really a must; see Sec.~\ref{deg}).  But the above algebra
of observables is too simple, and cannot by itself tell us
very much new.  It has to be extended.  Gour \cite{gour} has
shown how to introduce  a ``secret'' operator corresponding
to the quantum number $s$; his prescription, however,
presupposes a uniformly spaced spectrum for $\hat A$, which is
an idea to be tested here.  We, therefore, avoid including
something like this in the algebra.

In field theory of particles we would at this point introduce
fields and analyze them into creation and anhilation
operators; but ``field'' is an inappropriate concept for
black holes which are not even approximately pointlike
objects.  Nobody stops us, however, from defining the black
hole vacuum $|{\rm vac} \rangle$ (spacetime with no black
holes or particles of any sort), and creation operators for
black holes $\hat R_{njmqs}$ with the property that 
$|njmqs\rangle=\hat R_{njmqs}|{\rm vac} \rangle$.    This is essentially a
tautology, not a physical assumption, because we introduce as
many operators as there are states.  In contrast to field
theory, we do {\it not\/} assume that $\hat R_{njmqs}{}^N$
creates a state with $N$ black holes. 

By commuting $\hat A,  \hat {\bf J}$ and $\hat Q$ with the
$\hat R_{njmqs}$ and iterating we can make more operators. 
If this process continues indefinitely, no information can be
obtained from the algebra unless additional assumptions are
made. So let us suppose that the mentioned operators,
together with the unit operator $\hat I$, form a closed,
linear, infinite dimensional algebra.  This assumption has
two new features: the closure at some low level of
commutation, essentially a plea for simplicity, and the
linear character of the algebra when formulated in terms of
$\hat A$.  As we shall see in Sec.~\ref{spect}, this last
implies the additivity of horizon area, which is a reasonable
property.  By contrast, additivity of mass for two black
holes is not reasonable (nonlinearity of gravity), and this
is really the reason why one cannot assume linearity of the
algebra of
$\hat M$,
$\hat Q$,
$\hat {\bf J}$ and $\hat R_{njmqs}$.  In this sense $\hat A$
is special among all functions of the other black hole
observables. 

By definition $\hat R_{njmqs}{|{\rm vac}\rangle}$ is an
eigenstate of $\hat Q$ with eigenvalue
$qe$, so
\begin{equation}
\exp(\imath\chi \hat Q)\, \hat R_{njmqs}{|{\rm vac}\rangle} =
\exp(\imath\chi q e)\,
\hat R_{njmqs} {|{\rm vac}\rangle}
\label{charge}
\end{equation} 
for real $\chi$.  Thus, as already mentioned, $\hat Q$ is the
generator of (global) quantum gauge transformations, and this
is equivalent to requiring
\begin{equation}
\exp(\imath\chi \hat Q)\, \hat R_{njmqs} \exp(-\imath\chi
\hat Q) = 
\exp(\imath\chi q e)\,\hat R_{njmqs}
\label{gauge}
\end{equation} 
(check by operating with this  on ${|{\rm
vac}\rangle}$, recalling that $\hat Q{|{\rm vac}\rangle}=0$,
and recovering Eq.~(\ref{charge})).  Now expansion of
Eq.~(\ref{gauge}) to
$O(\chi)$ gives one of our essential commutators,
\begin{equation} 
[\hat Q, \hat R_{njmqs}] = q e\, \hat
R_{njmqs}.
\label{commuteQ}
\end{equation}

Now obviously
\begin{equation}
\exp(\imath\chi \hat J_z/\hbar)\, \hat R_{njmqs}{|{\rm
vac}\rangle} =
\exp(\imath\chi m)\, \hat R_{njmqs} {|{\rm vac}\rangle},
\label{eigenvalueJz}
\end{equation} 
Repeating the previous sort of argument we get
\begin{equation} 
[\hat J_z, \hat R_{njmqs}] = m\,\hbar\,
\hat  R_{njmqs}.
\label{commuteJz}
\end{equation} 
There is a more significant way to get
Eq.~(\ref{commuteJz}).  Since
$\hat R_{njmqs}|{\rm vac}\rangle$ is defined as a state with
spin quantum numbers $j$ and $m$, the collection of such
states with fixed
$j$ and all allowed $m$ must transform among themselves under
rotations of the black hole like the spherical harmonics
$Y_{j\mu}$ (or the corresponding spinorial harmonic when $j$
is half--integer).  But  ${|{\rm vac}\rangle}$ must obviously
be invariant under all rotations, so $\hat R_{njmqs}$ may be
taken to behave like an irreducible spherical tensor operator
of rank $j$ with the usual $2j+1$ components labeled by $m$
\cite{merzbacher}.  This gives Eq.~(\ref{commuteJz})
immediately.  In addition, defining the usual raising and
lowering operators, $\hat J_\pm\equiv J_x\pm \imath J_y$, we
infer
\begin{equation} 
[\hat J_\pm, \hat R_{njmqs}] =
\sqrt{j(j+1)-m(m\pm 1)}\,\hbar\,\hat R_{nj,m\pm 1,qs}.
\label{commuteJ+}
\end{equation}

We can use this and the identity
\cite{merzbacher} $\hat {\bf J}^2 = (1/2)(\hat J_+\hat J_- +
\hat J_-\hat J_+) +\hat J_z^2$ to work out $[\hat {\bf J}^2, 
\hat R_{\kappa\,
s}]$.  It has a rather complicated form; its first term is
$j(j+1)\hbar^2 \hat R_{njmqs}$ which is followed by two terms
having on their right hand sides $\hat J_+$ and $\hat J_-$,
respectively.  Thus operating with $[\hat {\bf J}^2, \hat
R_{njmqs}]$ on
${|{\rm vac}\rangle}$ and taking into account that $\hat {\bf
J} {|{\rm vac}\rangle}=0$ we get
\begin{equation}
\hat {\bf J}^2\,\hat R_{njmqs}{|{\rm vac}\rangle}\, =
j(j+1)\hbar^2 \hat R_{njmqs}{|{\rm vac}\rangle}
\label{eigenvalueJ^2}
\end{equation} 
This corresponds to the definition of $\hat
R_{njmqs}$ as creation operator of a black hole with angular
momentum quantum numbers $j$ and $m$. 

\subsection{Including $\hat A$ in the algebra}
\label{incl}

So far all we know about $\hat A$ is that it commutes with
$\hat {\bf J}$ and
$\hat Q$.  We can extend the algebra to it by using the
Jacobi {\it identity\/}
\begin{equation} 
[\hat W, [\hat V, \hat U]] + [\hat V, [\hat
U, \hat W]] + [\hat U, [\hat W,
\hat V]]=0.
\label{Jacobi}
\end{equation} 
Replacing $\hat W\rightarrow \hat A$, $\hat
U\rightarrow \hat R_{njmqs}$, and $\hat V$ in turn by $\hat
J_z$,  $\hat J_\pm$ and
$\hat Q$, and using Eqs.~(\ref{commuteQ}), (\ref{commuteJz})
and (\ref{commuteJ+}) as well as the mutual commutativity of
all observables, we obtain 
\begin{eqnarray}
 & & [\hat Q, [\hat A, \hat R_{njmqs}]]  =  q e\, [\hat A,
\hat R_{njmqs}].
\label{Q}
\\ & & [\hat J_z, [\hat A, \hat R_{njmqs}]  =  m\,\hbar\,
[\hat A,  \hat R_{njmqs}], 
\label{Jz}
 \\
 & & [\hat J_\pm, [\hat A, \hat R_{njmqs}]]  = 
\sqrt{j(j+1)-m(m\pm 1)}\,\hbar\,[\hat A,\hat R_{nj,m\pm,qs}]. 
\label{J+}
\end{eqnarray} The fact that these commutators mimic those in
Eqs.~(\ref{commuteQ}), (\ref{commuteJz}) and
(\ref{commuteJ+}) properly reflects the rotational and gauge
invariant status of $\hat A$ which forces $[\hat A, \hat
R_{njmqs}]$ to transform exactly like $\hat R_{njmqs}$.  
Note that all the previous commutation relations are
invariant under the redefinition $\hat A\rightarrow
\hat A+{\rm const.}$  We single out the $\hat A$ of physical
interest by the requirement that $\hat A\,{|{\rm
vac}\rangle}=0$.

According to our closure assumption, $[\hat A, \hat
R_{njmqs}]$ has to be a linear combination of some of the
operators $\hat I, \hat R_{n'j'm'q's'},
\hat A, \hat Q, \hat J_z$ and $\hat J_\pm$.  From
Eq.~(\ref{gauge}) and its analog for $\hat J_z$ and rotations,
it is clear that $\hat I, \hat Q$ and
$\hat A$, all of them rotational scalars and gauge
invariants, can only show up in the linear combination in the
cases $\{nqjms\} =\{n000s\}$.  Further, the triplet $\{\hat
J_{-1},\hat J_0,\hat J_{+1}\}\equiv \{\hat J_-,\hat J_z,\hat
J_+\}$ is gauge invariant and a spherical irreducible tensor
of rank one \cite{merzbacher}, so $\hat J_\nu$  with
$\nu=0, \pm 1$ can only show up in the linear combination in
the cases $\{nqjms\} =\{n01\nu s\}$.  Furthermore, by
rotational invariance all three $\hat J_\nu$ must occur with
like coefficient.  Finally, $\hat R_{n'j'm'q's'}$ can only
show up if its subscripts
$q'$, $j'$ and $m'$ match those of $[\hat A, \hat R_{njmqs}]$
(we cannot have another $j$ appearing since Eq.~(\ref{J+})
makes it clear that $[\hat A, \hat R_{njmqs}] $ contains a
single $j$).  In equations
\begin{equation} 
[\hat A, \hat R_{njmqs}]=\sum_{n's'}
h_{ns}^{n's'}\, \hat R_{n'jmqs'} 
+\delta_{q}{}^0\big[\delta_{j}{}^0\, (C_{ns}\hat I +
D_{ns}\hat Q + E_{ns}\hat A) + \delta_{j}{}^1\, F_{ns}\hat
J_{m}\big]
\label{prelim}
\end{equation}  
where $C_{ns}, D_{ns}, E_{ns}$ and $F_{ns}$
are c--numbers and $h_{ns}^{n's'}$ is a c--number matrix.

Let us now operate with Eq.~(\ref{prelim}) on the vacuum;
$\hat Q, \hat {\bf J}$ and $\hat A$ all anhilate it, so we are
left with
\begin{equation} 
a_n\, \hat R_{njmqs} {|{\rm vac}\rangle}
=\sum_{n's'} h_{ns}^{n's'}\, \hat R_{n'jmqs'} {|{\rm
vac}\rangle} +  \delta_{q}{}^0\,\delta_{j}{}^0\,C_{ns}{|{\rm
vac}\rangle}
\end{equation} 
However, it is clear that the states
$\{|nqjms\rangle\}$ are orthogonal to one another
(automatically in the space spanned by $n, q, j, m$ and by
Schmidt orthogonalization with respect to the $s$ quantum
number), and all of them to the vacuum.  The previous
equation will contradict this unless we demand $C_{ns}=0$
and  $h_{ns}^{n's'}=a_n\,\delta_n{}^{n'} \delta_s{}^{s'}$. 
In that case
\begin{equation} 
[\hat A, \hat R_{njmqs}] =a_n\, \hat
R_{njmqs} +\delta_{q}{}^0\big[\delta_{j}{}^0\, (D_{ns}\hat Q
+ E_{ns}\hat A) +
\delta_{j}{}^1\ F_{ns}\hat J_{m}\ \big].
\label{acomm} 
\end{equation} 

At this stage we exploit the freedom left
in $\hat R_{njmqs}$ to define a new set of creation operators
\begin{equation}
\hat {\cal R}_{njmqs}\equiv \hat R_{njmqs} +  (a_n)^{-1}
\delta_{q}{}^0\big[\delta_{j}{}^0\, (D_{ns}\hat Q +
E_{ns}\hat A) +
\delta_{j}{}^1\, F_{ns}\hat J_{m}\ \big]
\label{newR}
\end{equation} After the redefinition the algebra of  $
\{\hat I, \hat A, \hat Q, {\bf J},  \hat {\cal R}_{njmqs}\}$
is still closed, and the  $\hat {\cal R}_{njmqs}$ create
exactly the same states as the $\hat R_{njmqs}$. The
redefinition transforms Eqs.~(\ref{acomm}) and 
(\ref{commuteQ}) into
\begin{eqnarray} & &[\hat A, \hat {\cal R}_{njmqs}] =a_n\,
\hat {\cal R}_{njmqs}
\label{commuteA}
\\ & & [\hat Q, \hat {\cal R}_{njmqs}] =q\, \hat {\cal
R}_{njmqs},
\label{commuteQnew}
\end{eqnarray} 
but changes the forms of (\ref{commuteJz}) and
(\ref{commuteJ+}) slightly
\cite{ann_arbor}.

\subsection{The area spectrum}
\label{spect}

What is the state $\hat {\cal
R}_{njmqs}|n'q'j'm's'\rangle=\hat {\cal R}_{njmqs}\, \hat
{\cal R}_{n'q'j'm's'} {|{\rm vac}\rangle}$ ?  In field theory
we would unhesitatingly identify it as a two--black hole
state.  Here it is different.  To explain why, it is useful to
denote the quantum numbers
$\{njmqs\}$ collectively by a Greek index, $\kappa$,
$\lambda$ or $\mu$, as the case may be.  Likewise,  we denote
the operators $\hat Q$ and $\hat A$ by the common symbol
$\hat X$, and the latter's eigenvalues by 
$\xi_\lambda$. From Eqs.~(\ref{commuteA})-(\ref{commuteQnew})
and the Jacobi identity (\ref{Jacobi}) we discover that 
\begin{equation} 
[\hat X, [\hat {\cal R}_{\kappa}, \hat {\cal
R}_{\lambda}]] = (\xi_\kappa+\xi_\lambda) [\hat {\cal
R}_{\kappa},\hat {\cal R}_{\lambda}].
\label{commutepair}
\end{equation} 

Now by the closure condition 
\begin{equation} 
[\hat {\cal R}_{\kappa}, \hat {\cal
R}_{\lambda}] =\sum_{\mu}
\varepsilon_{\kappa\lambda}^{\mu}\hat R_{\mu} + \cdots
\label{twoR}
\end{equation} 
where the ellipsis signifies some linear
combination of $\hat A, \hat J_{m}$ and
$\hat Q$, and $ \varepsilon_{\kappa\lambda}^{\mu}$ are
c-numbers (structure constants).  Substituting in
Eq.~(\ref{commutepair}) gives
\begin{equation}
\sum_{\mu} \varepsilon_{\kappa\lambda}^{\mu} [\hat X,{\cal
R}_\mu]= (\xi_\kappa+\xi_\lambda) (\sum_{\mu}
\varepsilon_{\kappa\lambda}^{\mu} \hat {\cal R}_\mu+\dots),
\end{equation} which in view of
Eqs.~(\ref{commuteA})-(\ref{commuteQnew}) excludes  the $\
\cdots\
$ terms, and also tells us that whenever 
$\varepsilon_{\kappa\lambda}^{\mu}\neq 0$,   
\begin{equation}
\xi_\mu\equiv \xi_\kappa+\xi_\lambda 
\label{additivity}
\end{equation} 
for both types of $\hat X$. We shall assume
that for given $\kappa$ and $\lambda$ at least one of the
$\varepsilon_{\kappa\lambda}^{\mu}\neq 0$ does not vanish
(see below for the interpretation). 

Operating on ${|{\rm vac}\rangle}$ with Eq.~(\ref{twoR})  one
gets
\begin{equation} 
[\hat {\cal R}_{\kappa}, \hat {\cal
R}_{\lambda}]{|{\rm vac}\rangle}\, = \,|\bullet\rangle
\label{one}
\end{equation} 
where $|\bullet\rangle$ stands for a {\it
one\/}--black hole state, a superposition of states
$\{njmqs\}$ which by virtue of Eq.~(\ref{commutepair}) all
have a common area eigenvalue
$a_\kappa+a_\lambda$ and a common charge
$q_\kappa+q_\lambda$.  This $|\bullet\rangle$ is obviously a
physical state (it complies with the charge superselection
rule); it may involve superpositions of different
$j$ and $m$.  

We conclude that
\begin{equation} 
|\Psi\rangle \equiv \hat {\cal
R}_{\kappa}\,\hat {\cal R}_{\lambda}|{\rm
vac}\rangle\,=\,{\scriptscriptstyle 1\over\scriptscriptstyle
2}\,|\bullet\bullet\,\rangle\, +\,{\scriptscriptstyle
1\over\scriptscriptstyle 2}\,|\bullet\rangle.
\label{two}
\end{equation} 
Here $|\bullet\bullet\,\rangle=({\cal
R}_\kappa {\cal R}_\lambda + {\cal R}_\lambda {\cal
R}_\kappa)\,|{\rm vac}\rangle$ is obviously a two-black hole
state symmetric under exchange of the
$\kappa$ and $\lambda$ sets of quantum numbers.  Thus trying
to add an extra black hole to the state $\{njmqs\}$ actually
creates a linear combination of a one- and a two-black hole
states.  This is reasonable since classically black holes can
merge, so there is some amplitude for the new black hole to
fuse with the original one.  This is the ultimate
justification for our  assumption that at least one of the
$\varepsilon_{\kappa\lambda}^\mu$ is nonzero.  

Now using Eqs.~(\ref{commuteA})-(\ref{commuteQnew}) twice we
find that 
\begin{equation}
\hat X\,|\Psi\rangle =\hat {\cal R}_{\kappa}(\hat
X+\xi_\kappa)\hat {\cal R}_{\lambda}{|{\rm vac}\rangle}
=(\xi_\kappa+\xi_{\lambda}) |\Psi\rangle
\label{doublecommute} 
\end{equation} 
Thus the states $|\Psi\rangle$ and
$|\bullet\bullet\,\rangle$ both have sharp area
$a_\kappa+a_\lambda$ and sharp charge
$q_\kappa+q_\lambda$, just like $|\bullet\rangle$: area, like
charge, is an additive quantity for two black holes.   This
additivity  jibes with our geometric notion that areas of
separate objects are additive, and serves as further
justification for the assumptions leading to our algebra. 
And because
$|\Psi\rangle$ involves $|\bullet\rangle$,  we find, in
addition, that {\it the sum of two eigenvalues of $\hat Q$ or
$\hat A$ for a single black hole is also a possible
eigenvalue of $\hat
Q$ or $\hat A$, respectively, of a single black hole\/}.

It turns out that differences of eigenvalues of the $\hat X$
are also eigenvalues for one black hole.  Consider the
hermitian conjugates of
Eqs.~(\ref{commuteA})-(\ref{commuteQnew}),
\begin{equation} 
[\hat X, \hat {\cal
R}_{\kappa}^\dagger]=-\xi_\kappa \hat {\cal
R}_{\kappa}^\dagger.
\label{hermitian}
\end{equation} 
What is the meaning of $|\chi\rangle\equiv
\hat {\cal R}_{\kappa}^\dagger\, {|{\rm vac}\rangle}$ ? 
Operating with Eq.~(\ref{hermitian}) with $\hat X=\hat A$ on
${|{\rm vac}\rangle}$ and taking the scalar product with
$\langle \chi|$ shows that $\hat A$ would have a {\it
negative} average in the state
$|\chi\rangle$, unless this last state vanishes.  Thus, since
$\hat A$ is a positive definite operator,  $\hat {\cal
R}_{\kappa}^\dagger$ must anhilate the vacuum.  It seems very
plausible then that $|\Xi\rangle\equiv  \hat {\cal
R}_{\kappa}^\dagger \hat {\cal R}_{\lambda}\, {|{\rm
vac}\rangle}$ can only be a one--black hole state,
pressumably distinct from
$\hat {\cal R}_{\lambda}\, {|{\rm vac}\rangle}$.  Applying
Eqs.~(\ref{commuteA}),(\ref{commuteQnew}) and
(\ref{hermitian}) gives
\begin{equation}
\hat X\, |\Xi\rangle=\, \left(\hat {\cal
R}_{\kappa}^\dagger\hat X - \xi_\kappa
\hat {\cal R}_{\kappa}^\dagger
\right)\hat {\cal R}_{\lambda} {|{\rm vac}\rangle}\, =
\left(\xi_{\lambda}-\xi_\kappa\right)|\Xi\rangle.
\label{commuteconj}
\end{equation} 
This verifies our claim, with the obvious
caveat that if
$a_{\lambda} < a_\kappa$,  $\hat {\cal R}_{\kappa}^\dagger$,
must anhilate the state $\hat {\cal R}_{\lambda}\, {|{\rm
vac}\rangle}$ because negative black hole areas are
unacceptable (the same must happen if $a_{\lambda} =
a_\kappa$  with $q_{\lambda}\neq q_\kappa$ because a zero
area state is necessarily the vacuum, which bears no charge. 
Hence {\it the positive difference of two eigenvalues of
$\hat A$ for a single black hole is also a possible eigenvalue
of $\hat A$ of a single black hole; the difference of two
$\hat Q$ eigenvalues is a possible $\hat Q$ eigenvalue for
one black hole\/}.

We take it from Sec.~\ref{AI} that the spectrum of $\hat A$
is discrete.  Now the only discrete set of positive real
numbers that is unchanged under addition or absolute value
substraction of two members is the set of all the
natural numbers multiplied by some common factor. 
We conclude that the one-black hole area spectrum is just
$\{na_1|\, n=1, 2, \cdots\ \}$, where $a_1$ is some positive
scale of area (here we make no attempt to determine which
eigenvalues correspond to which charges and spin; see
\cite{ann_arbor}).  It is also clear that the two rules (in
italics) lead to a one-black hole charge spectrum composed of
all integers  multiplied by a common factor, $e$.  We
have thus formally obtained from the algebra the kind of area
spectra predicted long ago \cite{bek74}, and summarized in
Sec.~\ref{AI}.  For nonrotating neutral black holes, the
corresponding mass eigenvalues are proportional to $\{\surd
n\, |\, n=1,2, \dots\, \}$.    

Based on a variety of interpretations of the
nonrotating neutral black hole in canonical quantum gravity,
Schiffer (whose paper's title is used as title of the present
lecture), Peleg, Kastrup, Louko and M\"akel\"a, Barvinsky and
Kunstatter, Berezin, Vaz and Witten \cite{many} have all
obtained a mass spectrum of the mentioned form, but with no
consensus as to the exact numerical coefficient.  Some such
calculations give a not uniformly spaced spectrum
\cite{others}. Regarding RN black holes, M\"akel\"a and
Repo{\cite{makela_repo97} find the {\it sum\/} of areas of
outer and inner horizons to scale like an integer, and a
recent paper of theirs and coworkers extends this rule to
charged rotating black holes
\cite{makela_repo00}.  Vaz and Witten
\cite{vaz_witten00} find a law of this form rather for the
{\it difference\/} of these areas.  And Barvinsky, Das and
Kunstatter \cite{bar_das_kunst} find an {\it external\/}
horizon area with a spectrum precisely equally spaced, but
with its zero point shifted by a charge dependent quantity.
Overall, the predictions are thus similar to those following
from the algebra.

\subsection{Degeneracy of area eigenvalues}
\label{deg}

According to Mukhanov \cite{mukh86} degeneracy plays a
central role in any discussion of area levels.  What can we
say about it from our approach ?  By rotational invariance
neither area eigenvalues nor degeneracy factors can depend on
the quantum number $m$.  Let us assume that the spectrum
$a_n=na_1$ for $n=1,2,\,\cdots\,$ is common to every
combination of quantum numbers $j$ and $q$ (there is one
alternative to this \cite{ann_arbor}).  The degeneracy factor
will be of the form $g_n=g_n(j,q)$. Now for fixed
$\{n_\kappa, j_\kappa, m_\kappa, q_\kappa\}$ with not all of
$j_\kappa, m_\kappa$ and $q_\kappa$ vanishing, there are
$g_{n_\kappa}(j_\kappa,q_\kappa)$ independent one--black hole
states $\hat {\cal R}_\kappa{|{\rm vac}\rangle}$
distinguished by the values of $s$.  Analogously, the set
$\{n_\lambda=1, j_\lambda=0, m_\lambda=0, q_\lambda=0\}$
specifies $g_1(0,0)$ independent states $\hat R_\lambda{|{\rm
vac}\rangle}$, all different from the previous ones because
not all quantum numbers agree.  One can thus form
$g_1(0,0)\cdot g_{n_\kappa}(j_\kappa,q_\kappa)$ one--black
hole states, $[\hat R_\kappa, \hat R_\lambda]{|{\rm
vac}\rangle}$, with area eigenvalues
$(n_\kappa + 1)a_1$, and charge and spin just like the states
$\hat R_\kappa{|{\rm vac}\rangle}$.  Let us assume these new
states are independent.  Then their number cannot exceed the
total number of states with area $(n_\kappa + 1)a_1$ and
quantum numbers $j_\kappa$ and $q_\kappa$:
$g_{{n_\kappa} +1}(j_\kappa,q_\kappa)
\geq g_1(0,0)\cdot g_{n_\kappa}(j_\kappa,q_\kappa)$. 
Iterating this inequality starting from
$n_\kappa=1$ gives (we drop  $\kappa$)
\begin{equation} 
g_n(j,q)\geq  g_1 (j,q)\cdot g_1(0,0)^{n-1}.
\label{degeneracy2}
\end{equation}  

If $g_1(0,0)\neq 1$, Eq.~(\ref{degeneracy2})  tells us that
the degeneracy rises at least exponentially with area.  Since
the area spectrum is rather sparse, black hole entropy must
receive its principal contribution from the logarithm of the
degeneracy of area levels, so what we have just found is that
black hole entropy must grow at least as fast as the horizon
area.  Thus we have a pleasant microscopic explanation of the
black hole entropy--area relation.  Further, as first
emphasized by Mukhanov \cite{mukh86}, one can calibrate the
area spectrum by use of the degeneracy $\leftrightarrow $
entropy correspondence relation.  For $g_1(0,0)=2$ this gives
$a_1=4\hbar\ln 2$ with the corresponding mass spectrum
$\{ (\hbar\ln 2/4\pi)^{1/2}\surd n\,|\,n=1,2,\,\cdots\,\}$
\cite{bek_mukh}.

\noindent
{\bf Acknowledgments\hskip 12pt} This research is supported by
grant No. 129/00-1 of the Israel Science Foundation.

\end{document}